\documentclass[11pt]{amsart}
\usepackage[centertags]{amsmath}
\usepackage{amsfonts}
\usepackage{amssymb}
\usepackage{amsthm}
\usepackage[english]{babel}
\usepackage[latin2]{inputenc}
\usepackage{graphicx}

\title[Comparation the HC and its environment]{Comparation based bottom-up and top-down filtering model of the hippocampus and its environment}

\author[A. L\H{o}rincz]{Andr\'{a}s L\H{o}rincz}

\thanks{\noindent \newline Department of Information Systems \\ E\"otv\"os Lor\'and University \\ P\'azm\'any P\'eter s\'et\'any 1/C \\
Budapest, Hungary H-1117 \\ email: lorincz@inf.elte.hu}

\begin{document}

 \maketitle

\begin{abstract}
Two rate code models -- a reconstruction network model and a control model -- of the hippocampal-entorhinal loop are
merged. The hippocampal-entorhinal loop plays a double role in the unified model, it is part of a reconstruction
network and a controller, too. This double role turns the bottom-up information flow into top-down control like
signals. The role of bottom-up filtering is information maximization, noise filtering, temporal integration and
prediction, whereas the role of top-down filtering is emphasizing, i.e., highlighting or `paving of the way' as well as
context based pattern completion. In the joined model, the control task is performed by cortical areas, whereas
reconstruction networks can be found between cortical areas. While the controller is highly non-linear, the
reconstruction network is an almost linear architecture, which is optimized for noise estimation and noise filtering. A
conjecture of the reconstruction network model -- that the long-term memory of the visual stream is the linear feedback
connections between neocortical areas -- is reinforced by the joined model. Falsifying predictions are presented; some
of them have recent experimental support. Connections to attention and to awareness are made.
\end{abstract}
\clearpage \newpage
\section{Introduction}
\label{s:intro}

Ever since the discovery of the central role of the hippocampus and its adjacent areas in memory formation
\cite{sidman68some,milner72disorders}, numerous studies and models dealt with the properties and the possible functions
of the hippocampus and its environment. The number of new experimental findings is increasing and highlight the
complexity of the behavior of memory. Although views are strikingly different, they seem to have their own,
experimentally supported merits. The interested reader is referred to the literature for excellent reviews on the
hippocampus, e.g., \cite{squire92memory}, or \cite{hasselmo99neural} and \cite{oreilly99conjunctive}.  The majority of
the models have been developed to describe one part (mainly the CA3 field) of the hippocampus (see, e.g.
\cite{levy96sequence,kali00involvement} and references therein). Some recent models have made attempts to develop an
integrating model of the HC
\cite{rolls89functions,hasselmo96encoding,lisman99relating,eichenbaum00cortical,hasselmo02proposed} and see also the
collection of theoretical papers \cite{hippocampus96gluck}. It is known though, that hippocampus is deeply embedded in
the neocortical information flow through the entorhinal cortex (EC). This fact explains the emergence of a few EC-HC
models like \cite{mcclelland95why,myers95dissociation,rolls00hippocampo}. Embedding is justified in most of them. For
example, McClelland et al. (\cite{mcclelland95why}) emphasizes the necessity of a dual system for the seemingly
contradictory tasks of learning specifications and allowing for generalization.

The computational model that we present here, has its origin in the old standing proposal that the hippocampus and/or
its environment serve as a `comparator' \cite{grastyan59hippocampal,sokolov63higher,vinogradova75functional}. There are
more recent works along this subjects. Oftentimes models use somewhat different nomenclature, e.g., the focus is placed
on match/mismatch detection \cite{ranck73studies,okeefe78hippocampus,grossberg82processing}. Match/mismatch detection
is closely related to familarity/novelty detection, another direction of theoretical efforts to describe medial
temporal lobe areas \cite{otto92neuronal,rolls93responses,wiebe88dynamic}. Note that precise distinction between
orienting, salient and novel stimuli is not an easy matter \cite{rugg95cognitive}.

This work is about the merging two comparator  based models: the control model of the entorhinal-hippocampal loop
\cite{lorincz98forming} and the reconstruction network model of the same loop \cite{lorincz00parahippocampal}. Both
models have their own merits. For example, there is a large body of experimental data supporting the idea that
attention shapes (influences, controls) perception. For excellent reviews, see, e.g.,
\cite{duncan99attention,posner00attention,laberge00networks} and references therein. The control model involves the
comparator function, because -- by construction -- control concerns the difference between desired and actual
parameters. On the other hand, the reconstruction network is also a comparator: it has a hidden layer, works as an
auto-associator, and compares input to the auto-associated \textit{reconstructed input}. It turns out that the
reconstruction network is an appealing structure for experience based optimization of noise filtering
\cite{lorincz02mystery}. We shall merge the two comparator structures and shall map the merged structure to the
entorhinal-hippocampal loop and its environment. This merging will enable us to make physiological predictions
concerning persistent activities, delay properties, long-term memory and statistic versus one-shot learning.

There are two approaches, which should be mentioned, because both of them find their place in the present model. Gluck
and Myers \cite{gluck93hippocampal} have designed a model to perform reconstruction \textit{and} classification
together for modeling some properties of the hippocampus. Rao and Ballard
\cite{rao97dynamic,rao99optimal,rao99predictive} have suggested an integrating model of the visual stream by exploring
a Kalman-filter analogy to cope with the input and system uncertainties (treated as noise) and presented a hierarchy
for error correction and prediction using top-down inference from higher levels. Kalman-filter is a kind of
reconstruction or generative network, which uses an internal representation to generate expected inputs. The mapping of
the proposed function onto the anatomical substrate has remained elusive. The Kalman-filter model, which is an
approximation to our model, has been criticized because such recurrent loop structures are slow for feedforward
processing found in neocortical areas \cite{koch99predicting}. Our model resolves this problem.

Mathematical theorems and numerical studies concerning individual components and combinations of those have been
presented elsewhere, or have been made available in the form of technical reports. For example, hierarchical
reconstruction networks \cite{lorincz02mystery}, the dynamics of the network as well as the order of learning in
reconstruction networks \cite{lorincz02relating}, implicit memory phenomena, such as priming \cite{lorincz02relating}
and category formation in reconstruction networks \cite{keri02categories} are provided in the cited papers. Numerical
studies concerning the control architecture can be found in
\cite{szepesvari97neurocontroller,szepesvari97approximate,SzLo98JRS}. Mathematical considerations as well as numerical
studies of the control architecture embedded into the reinforcement learning framework have been thoroughly described
in \cite{szita03epsilon}. Connections to reinforcement learning, the key to fast, possibly one-shot memory encoding
have been presented in \cite{Kalmar98Module-based,szita03epsilon} as well as in \cite{szita03kalman}. The mathematical
considerations have been complemented by some generalization concerning the control architecture in order to meet the
requirements posed by the merging of the two architectures. This slight mathematical generalization can be found in the
Appendix of a technical report \cite{lorincz_http_arxiv_CHF}. The novelty of the present work is in the merging of the
two models and in the description how the two models can be merged. We shall find that stability properties of the
controller are improved by the merging: The reconstruction network filters the input noise of the controller.
(Considerations on noise sensitivity of the controller can be found in \cite{szepesvari97approximate}.) Here, the model
is described \textit{by words} and \textit{in figures}. The interested reader may wish to consult the cited works about
the mathematical details.

In what follows, first, the terminology and the preliminaries are reviewed (Section \ref{s:notations}). In Section
\ref{s:model}, the merging of the control and reconstruction architectures into a single building block of a
hierarchical structure is described. The closure of the hierarchy provides the view that the hippocampus plays a double
role; it is part of a controller and contributes to a reconstruction network, too. Section \ref{s:discussion} deals
with the mapping of the control and reconstruction architectures to the entorhinal-hippocampal loop and its
environment. Physiological properties captured by the model as well as some falsifying predictions are listed and
explained in this section. Conclusions are drawn in Section \ref{s:conc}.

\section{Notations and preliminaries}
\label{s:notations}

\begin{figure}[h]
\centering
\includegraphics[width=85mm]{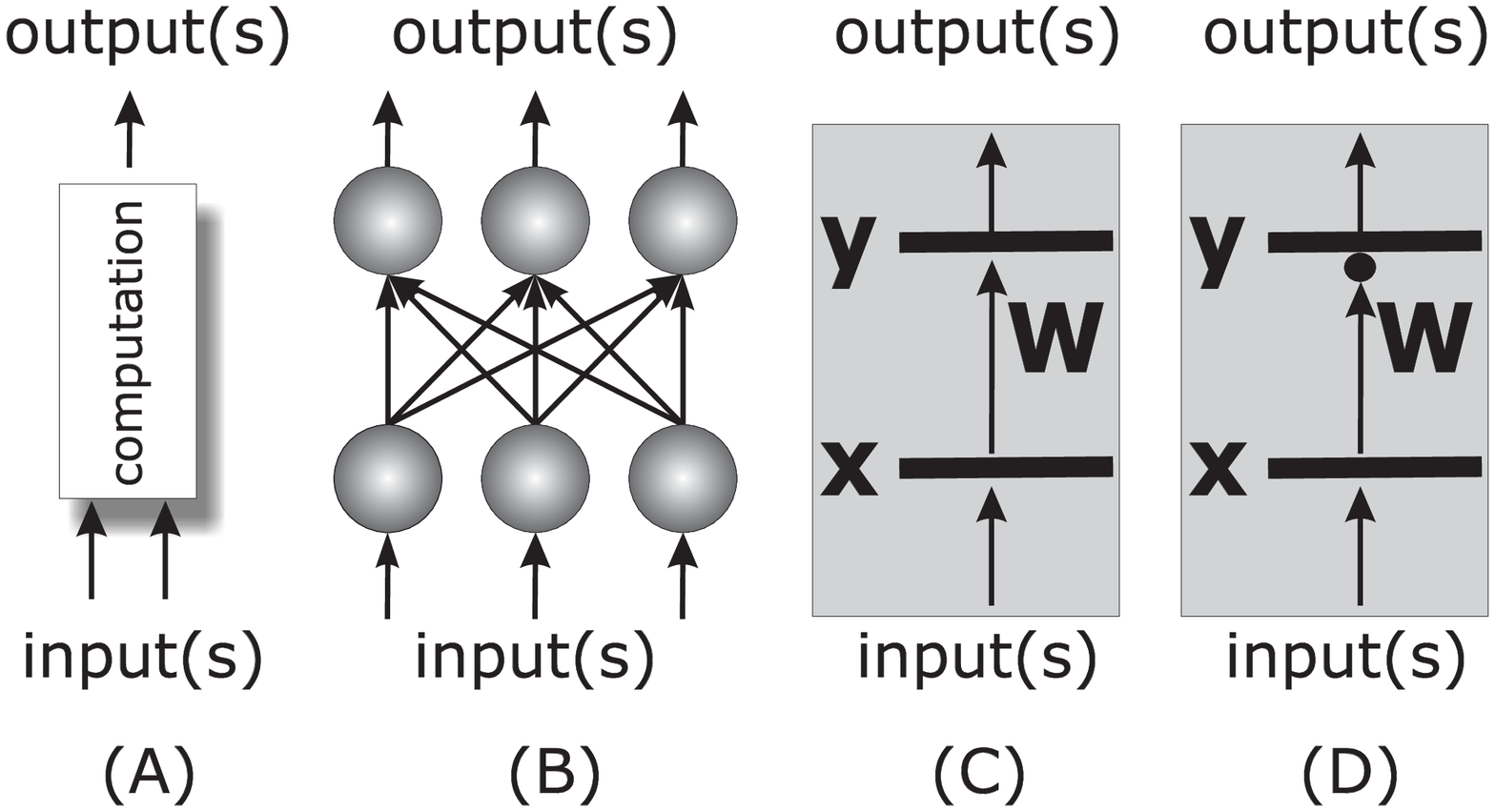}
 \caption{\textbf{Notations}
 \newline \noindent \textbf{A:} Control representation of input output
systems.
  Computations are performed in the box. \newline \textbf{B:} Neural
  network representation of computations: inputs are received by
  input neurons and are (non-linearly) transformed by connections
  and the output neurons, which provide the outputs.
  An output neuron could be an input neuron of the next processing
  stage. \newline \textbf{C:} Linear neural transformation. Input: $\mathbf{x}$,
  transformation $\mathbf{W}$, output: $\mathbf{y}$,
  $\mathbf{y}=\mathbf{W}\mathbf{x}$. \newline \textbf{D:} Non-linear neural transformation.
  $\mathbf{y}=f(\mathbf{W}\mathbf{x})$. Graphical form: arrow with a circle. More than one
  transformation may exist between layers. Recurrent network
  is a neural layer with a transformation that targets the same
  layer.
  \newline \textit{Terminology in the context of neurobiology:} \newline Layer is encompassed
  by an area (e.g., typical neocortical areas are made of 6 layers), or it is a subfield,
  such as the CA3 and CA1 regions of
  the hippocampus.
  Transformations may correspond to (i) excitatory synapses
  connecting layers or targeting neurons of the same layer, such
  as the \textit{recurrent collaterals} and the \textit{associative connections}
  of the CA3 subfield of the hippocampus and the intra-layer excitatory connections
  of layers II and III of the neocortex or (ii) inhibitory synapses
  between layers or within layers, such as the rich interneural networks
  in the hippocampus.}
\label{f:notat}
\end{figure}

Notations of the control field and notations of neuron networks differ. In control theory, input-output systems are
considered. In graphical form, box denotes the system and arrows  denote the system's input(s)  and output(s).
Processing occurs in the box (Fig.~\ref{f:notat}(A)). On the contrary, artificial neural networks consist of
computational units, the putative analogs of real neurons. The units, also called neurons, receive (provide) inputs
(outputs) through the connection structure, and this internal functioning is drawn explicitly. Neurons execute simple
computations, like summing up inputs, thresholding and alike. The main part of the neural network performs distributed
computation using the connection structure performing (non-linear) filtering. This distributed filter system, which may
connect all neurons, is called the connection system, weights, or synapses. In a neural network architecture, different
neural layers are distinguished. Connections between these layers are explicitly drawn in most cases
(Fig.~\ref{f:notat}(B)). Computations of neural networks between their inputs and outputs can be given in the following
condensed form:
\begin{equation}\label{eq:neur}
\mathbf{y} = f(\mathbf{Wx})
\end{equation}
where input and output  are denoted by $\mathbf{x} \in \mathbb{R}^n$ and $\mathbf{y} \in \mathbb{R}^m$, respectively,
linear transformation from $\mathbb{R}^n$ to $\mathbb{R}^m$ is represented by matrix $\mathbf{W} \in \mathbb{R}^{m
\times n}$, the connections, and function $f$ denotes a component-wise non-linearity. If this function is the identity
function, then we have a linear network. Here, a simplified notation will be used: neural layers will be denoted by
horizontal thick lines. Any particular set of connections between two layers will be represented by a single arrow. A
feedforward linear network is depicted in Fig.~\ref{f:notat}(C). The graphical form of a network with component-wise
non-linearity is shown in Fig.~\ref{f:notat}(D). Different transformations may exist between two layers.
\textit{Recurrent connections} (also called `recurrent collaterals') target the same layer where they originate from.
Equation~\ref{eq:neur} simplifies to the usual input-output mapping of a single neuron for $m=1$.

\subsection{Preliminaries} \label{ss:prelim}

\subsection*{Controller} \label{ss:control}

\begin{figure}
\centering
 \includegraphics[width=66mm]{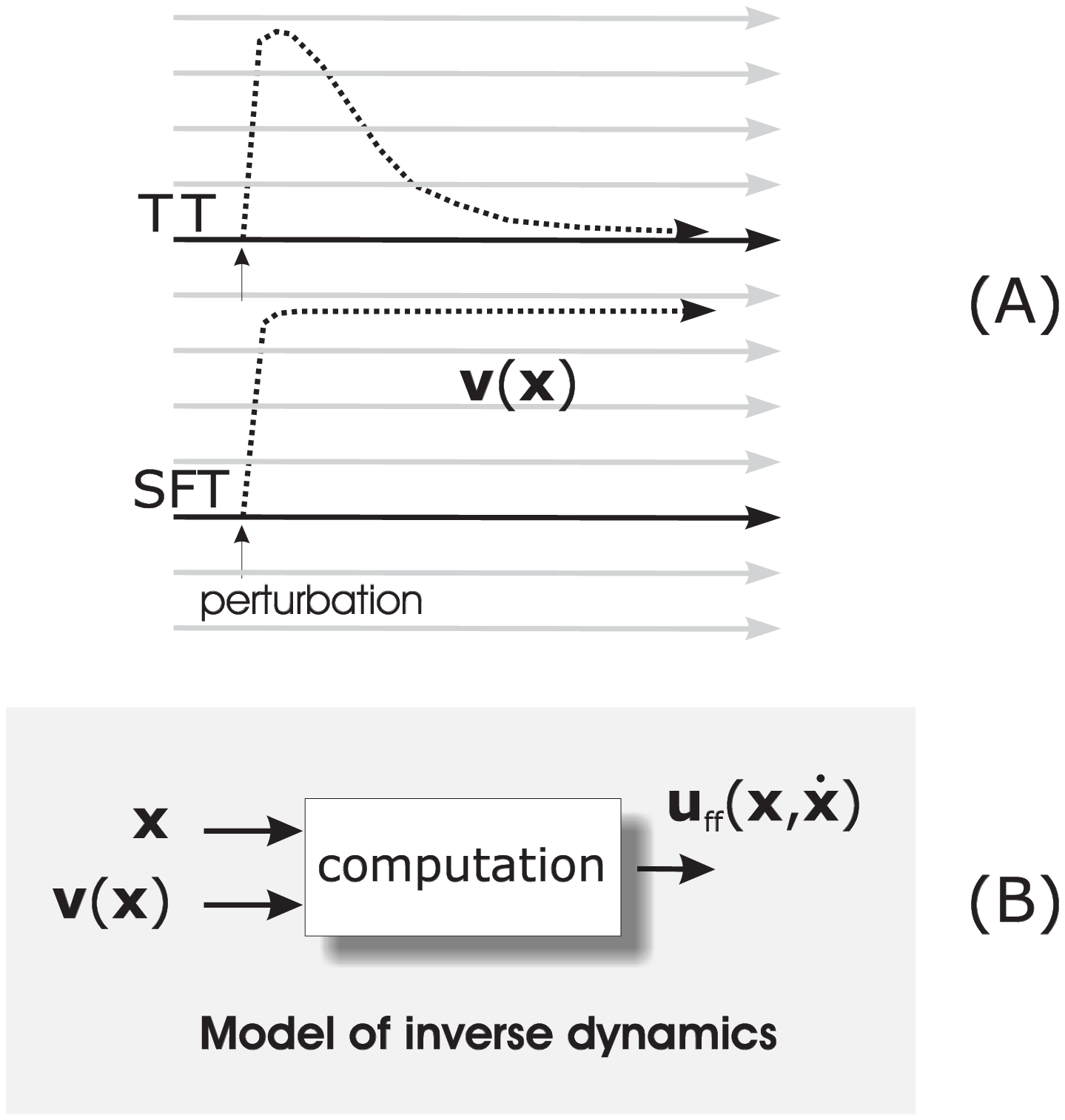}
  \caption{\textbf{Tracking and inverse dynamics)}
  \newline \noindent  \textbf{A:} \textbf{TT:} Horizontal lines represent different nearby trajectories in 2 dimensions.
  Black line: initial trajectory. Dotted line: perturbed trajectory. Upon perturbations the system (the `plant') should return
  to the original trajectory, which is also specified in time (not shown).
  \newline \textbf{SFT:} Horizontal lines represent a small homogeneous part of a
  speed-field $\mathbf{v}(\mathbf{x})$ to be tracked embedded into a 2 dimensional space $\mathbf{x}$ .
  Black line: initial speed trajectory. Dotted line: perturbed motion. Upon perturbation, the plant
  adjusts its speed to the prescribed speed in its actual neighborhood.\newline
  \textbf{B:} Inverse dynamics produces a control vector for a given pair of state and the belonging desired speed.}\label{f:SFT}
\end{figure}

Our control model  is formulated in terms of state dependent directions pointing towards target positions. A mapping
which renders direction (change of state, or change of state per unit time, i.e., velocity) to each state is called
speed-field. A particular speed-field is given, for example, by the difference vectors between the target state and all
other states. An important feature of speed-field is that motion is not specified in time. The control task is defined
as moving according to the speed-field at each state. This control task is called speed-field tracking (SFT). For a
review on SFT, see, e.g., \cite{HwaAh92}. The control task of path (also called trajectory) tracking is different from
SFT. This difference is illustrated in Fig.~\ref{f:SFT}. One might say that SFT is less stringent and puts more
emphasis on the global goal than on the tracking of a prescribed trajectory.

The dynamic equation  of a system is a (possibly continuous) set of differential equations. This set of equations
determines the change of state per unit time $\mathbf{\dot{x}}\approx \frac{\Delta \mathbf{x}}{\Delta t}$ given the
state of the plant (also called the system under control) $\mathbf{x} \in \mathbb{R}^n$ and the external forces acting
upon the system, including the control action $\mathbf{u} \in \mathbb{R}^p$. Let
\begin{equation}\label{eq:plant0}
\mathbf{\dot{x}} = \mathbf{f}(\mathbf{x},\mathbf{u})
\end{equation}
denote the dynamics of the system under control. This is what the system does.

Inverse dynamics works in the opposite way: given the state and (desired) change of state, inverse dynamics provides
the control vector. The controller, in turn, maps state and speed to control action. Let $\mathbf{v}(\mathbf{x}) \in
\mathbb{R}^n$ denote the \textit{desired} change of state (the desired momentum). Assume that we have an approximate
feedforward model of the inverse dynamics:
\begin{eqnarray}\label{eq:uff}
\mathbf{u}_{ff} &=& \mathbf{u}_{ff}(\mathbf{x},\mathbf{v}(\mathbf{x}))\\
&=&
\mathbf{A}(\mathbf{x})\mathbf{v}(\mathbf{x})+\mathbf{b}(\mathbf{x})\label{eq:uff2}
\end{eqnarray}
where $\mathbf{A}(\mathbf{x}) \in \mathbb{R}^{n \times n}$, $\mathbf{b}(\mathbf{x})\in \mathbb{R}^n$. Equation
\ref{eq:uff} represents an input-output system receiving the state and the desired momentum, which is denoted by
$\mathbf{v}(\mathbf{x})$ and providing output (the control vector $\mathbf{u}_{ff}:\mathbb{R}^{n \times n} \rightarrow
\mathbb{R}^p $). This control vector could be used directly to influence the plant and it is called \textit{feedforward
controller}. If the inverse dynamics is perfect then using this control vector directly, that is inserting this control
vector into the dynamic equation $\mathbf{\dot{x}} = \mathbf{f}(\mathbf{x},\mathbf{u})$, the desired change of state is
achieved: The perfect feedforward control vector $\mathbf{u}^*_{ff}$ makes the plant to produce momentum
$\mathbf{v}(\mathbf{x})$:
\begin{equation}\label{eq:perf_uff}
\mathbf{v}(\mathbf{x}) = \mathbf{f}(\mathbf{x},\mathbf{u}^*_{ff}(\mathbf{x},\mathbf{v}(\mathbf{x})))
\end{equation}

\begin{figure}
\centering
 \includegraphics[width=90mm]{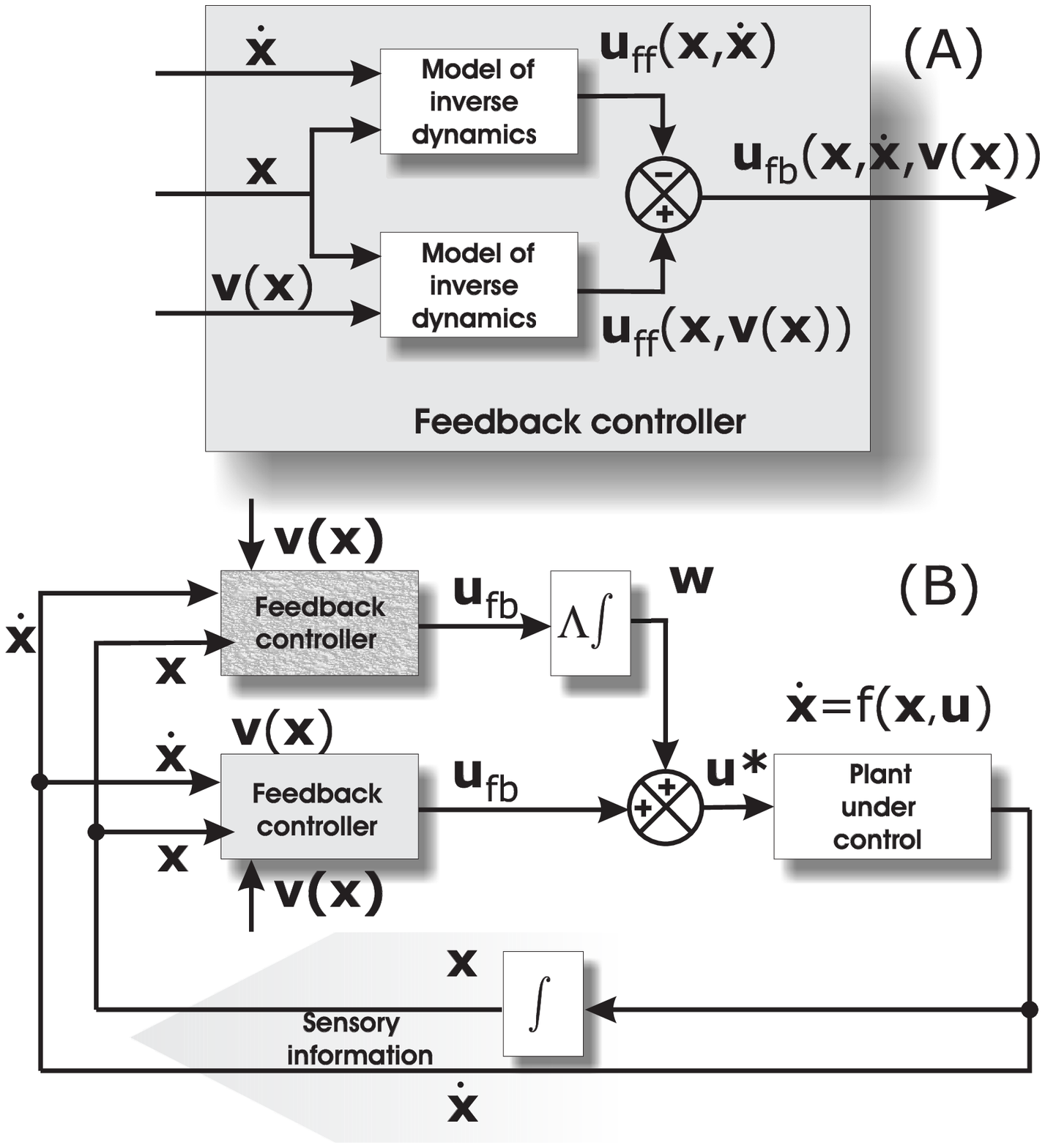}
  \caption{\textbf{Robust controller for speed-field tracking tasks}
  \newline \noindent \textbf{A:} The model of the inverse dynamics is inputted by the actual state $\mathbf{x}$ and the
  desired momentum (or desired speed vector) $\mathbf{v}(\mathbf{x})$.  The output of the model is the
  \textit{feedforward} control vector
  $\mathbf{u}_{ff}=\mathbf{u}_{ff}(\mathbf{x},\mathbf{v}(\mathbf{x}))$.
  The feedforward control vector may need corrections. The \textit{feedback} control vector is the
  difference between the feedforward control vector and the \textit{experienced control vector}
  $\mathbf{u}_{ff}(\mathbf{x},\mathbf{\dot{x}})$:   $\mathbf{u}_{fb}=
  \mathbf{u}_{ff}(\mathbf{x},\mathbf{v}(\mathbf{x}))-\mathbf{u}_{ff}(\mathbf{x},\mathbf{\dot{x}})$.
  \newline \textbf{B:} The static and dynamic state (SDS) feedback controller. The output of the feedback controller is (i) applied directly and
  (ii) it is integrated by time multiplied by the gain factor and the result is added to form
  the almost correct control vector $\mathbf{u}^{*}$.
  \textit{Differing texture} of the two feedback controllers denotes that computations
  in these input-output devices can differ as long as certain mathematical conditions concerning the sign
  of the components of the control vectors are met.}\label{f:robust_controller}
\end{figure}

If the feedforward  control vector is imprecise then error (a difference between the desired momentum and the
experienced momentum) $\mathbf{e}_c=\mathbf{v}(\mathbf{x})-\mathbf{\dot{x}}$ appears. To correct this error, the (same
or another) model of the inverse dynamics can be used. The error correcting controller is called \textit{feedback
controller} \cite{szepesvari97neurocontroller}: The output of the feedback controller $\mathbf{u}_{fb}$
\begin{equation}\label{eq:fb}
\mathbf{u}_{fb} = \mathbf{u}_{ff}(\mathbf{x},\mathbf{v}(\mathbf{x}))- \mathbf{u}_{ff}(\mathbf{x},\mathbf{\dot{x}})
\end{equation}
can be temporally integrated
\begin{equation}\label{eq:u}
\mathbf{w}(t) = \int_{-\infty}^t \mathbf{\dot{w}} dt = \int_{-\infty}^t \Lambda \mathbf{u}_{fb} \,dt
\end{equation}
and the two terms, together:
\begin{equation}\label{eq:u_}
\mathbf{u}^{*} = \mathbf{u}_{fb}+ \mathbf{w}
\end{equation}
provide an globally stable and almost perfect controller \cite{szepesvari97approximate,lorincz_http_arxiv_CHF}. This
scheme is  depicted in Fig.~\ref{f:robust_controller}. Note that Eq.~\ref{eq:uff2} allows one to write
$\mathbf{u}_{fb}$ as
\begin{equation}\label{eq:u_1}
\mathbf{u}_{fb}=\mathbf{A}(\mathbf{x})(\mathbf{v}(\mathbf{x})-\mathbf{\dot{x}})
\end{equation}
The controller allows for temporal changes of the non-linear terms $\mathbf{A}(\mathbf{x})$ and
$\mathbf{b}(\mathbf{x})$, a rare property in the control literature. The condition of this property make the working of
the architecture strongly non-linear \cite{lorincz01ockham} and will be discussed later. Also, the two feedback
controllers of Fig.~\ref{f:robust_controller}(B) can differ \cite{lorincz_http_arxiv_CHF}.
\begin{eqnarray}\label{eq:u_diff}
\mathbf{\dot{w}} &=& \Lambda \mathbf{A}(\mathbf{x})(\mathbf{v}(\mathbf{x})-\mathbf{\dot{x}}) \\
\mathbf{u}^{*} &=&  \mathbf{\hat{A}}(\mathbf{x})(\mathbf{v}(\mathbf{x})-\mathbf{\dot{x}})+ \mathbf{w}
\end{eqnarray}
where $\mathbf{A}$ can be different from $\mathbf{\hat{A}}$. Both controllers operate by \textit{comparing} `desired'
and `experienced' quantities. Under certain conditions, global stability is reached by the two controllers. The proof
relies on an extension to Ljapunov's second method
\cite{szepesvari97neurocontroller,szepesvari97approximate,szepesvari97robust,lorincz_http_arxiv_CHF}. The controller
architecture of Fig.~\ref{f:robust_controller} is called the static and dynamic state feedback controller, or SDS
controller, for short.

\subsection*{Reconstruction network} \label{ss:recnet}

\subsection*{Basic reconstruction network} \label{ss:basic_recnet}

\begin{figure}
\centering
 \includegraphics[width=88mm]{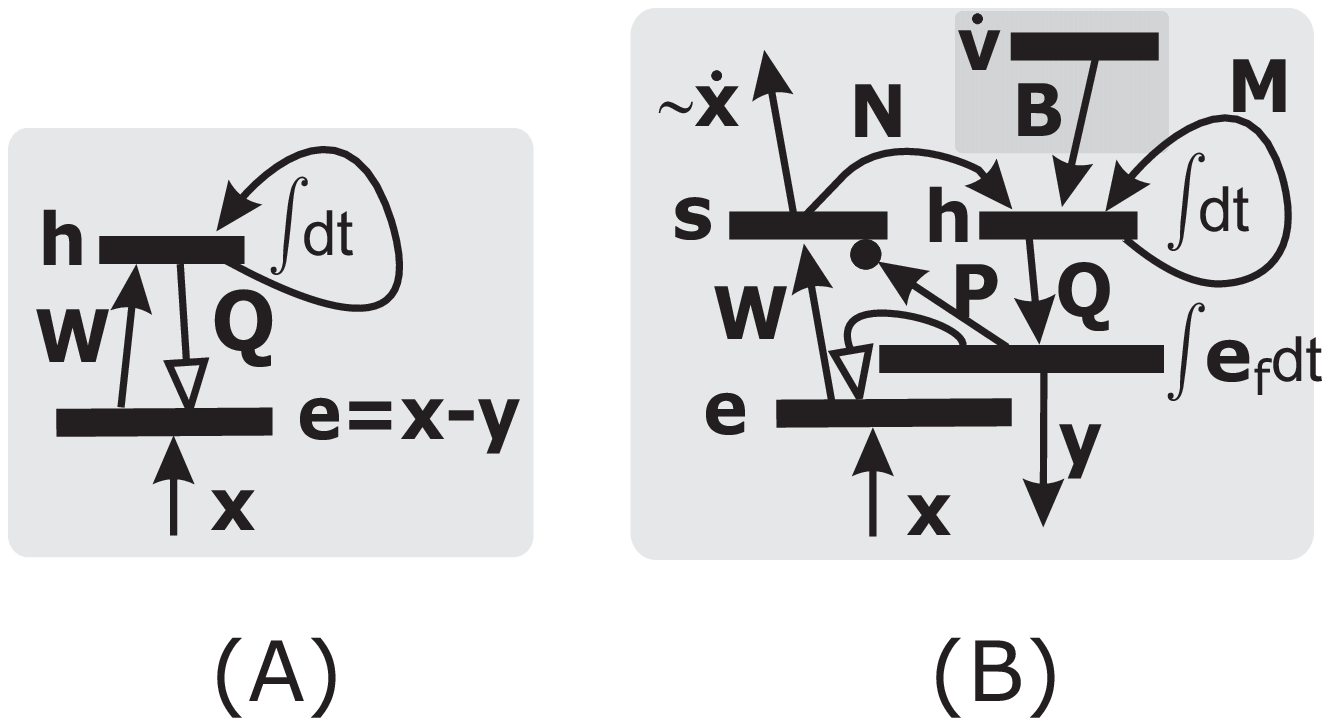}
  \caption{\textbf{Reconstruction networks}
\newline   \noindent  \textbf{A:} Simplest reconstruction network.
  $\mathbf{x}$ and $\mathbf{y}$: input and reconstructed input, respectively,
  $\mathbf{W}$ and $\mathbf{Q}$: bottom-up (BU) and top-down (TD) transformations, respectively.
  $\int dt$: summation in (or integration by) time.
  \newline
  \textbf{B:} Reconstruction network capable of noise filtering and pattern completion.
  $\mathbf{s}$: BU processed reconstruction error with maximized information,
  $\mathbf{h}$: activity vector of the hidden (model) layer,
  $\mathbf{N}$: transformation from BU processed reconstruction error layer to the model
  layer.
  $\mathbf{P}$: linear BU transformation followed by sparsifying thresholding. Thresholding gates
  components that deliver noise, but no information.
  For a perfectly tuned network, $\mathbf{P}=\mathbf{W}$, and either
  $\mathbf{QNW}=\mathbf{I}$ ($s \leq r$) or $\mathbf{NWQ}=\mathbf{I}$ ($r \leq s$).
  $\mathbf{M}$: recurrent collateral system for temporal integration and pattern completion.
  Dark gray inset: $\mathbf{B}$ and $\mathbf{\dot{v}}$ are tools of control. The controller provides the temporal
  derivative of the control vector $\mathbf{B}\mathbf{v}$, where $\mathbf{v}$ is the desired momentum of
  the controller. The derivative $\mathbf{B}\mathbf{\dot{v}}$ is integrated at the internal representation.}\label{f:recnets}
\end{figure}

The basic reconstruction network (Fig.~\ref{f:recnets}(A)) has two layers: the reconstruction error layer and the
hidden layer. The reconstruction error computes the difference ($\mathbf{e} \in \mathbb{R}^r$) between input
($\mathbf{x} \in \mathbb{R}^r$) and reconstructed input ($\mathbf{y} \in \mathbb{R}^r$):
$\mathbf{e}=\mathbf{x}-\mathbf{y}$. Reconstructed input $\mathbf{y}$ is produced by the hidden (internal)
representation $\mathbf{h} \in \mathbb{R}^s$ via top-down transformation $\mathbf{Q}$ where $\mathbf{Q}\in
\mathbb{R}^{r \times s}$ and $r \leq s$ or $r>s$ are both possible at the expense of some mild non-linearities (see,
e.g., \cite{OlFi97} and references therein). The hidden representation is corrected by the bottom-up transformed form
of the reconstruction error $\mathbf{e}$ that is by $\mathbf{We}$ (where $\mathbf{W}\in \mathbb{R}^{s \times r}$). The
process of correction means that the previous value of the hidden representation is to be maintained and the correcting
amount needs to be added. In turn, the hidden representation has a self-excitatory connection set, which maintains the
activities. Correction occurs in time and this (continuous or discrete) temporal collection of correcting terms is
denoted by $\int dt$. For sustained input $\mathbf{x}$ the iteration will stop when $\mathbf{Qh}=\mathbf{x}$ or when
the bottom-up transformed values of $\mathbf{Qh}$ and $\mathbf{x}$ are equal: $\mathbf{WQh}=\mathbf{Wx}$. In turn, for
each input, the hidden representation is determined by top-down transformation $\mathbf{Q}$. The bottom-up
transformation can restrict the range of the reconstruction. For example, if $\mathbf{W}$ projects to a subspace then
reconstruction can be executed only within that subspace.

For $s \leq r$, we call the network perfectly tuned if $\mathbf{W}=(\mathbf{Q}^T\mathbf{Q})^{-1}\mathbf{Q}^T$, i.e., if
$\mathbf{WQ}=\mathbf{I}$ ($\mathbf{I} \in \mathbb{R}^{s \times s}$). In this case, activities of the hidden layer
become perfect after a single bottom-up processing step and the network works alike to feedforward nets.

\subsection*{Reconstruction network plus} \label{ss:full_recnet}

The simple network of Fig.~\ref{f:recnets}(A) can be extended to support noise filtering. To this purpose, the
reconstructed input vector $\mathbf{y}$ is represented separately. Input vector $\mathbf{x}$ is compared to the
reconstructed input $\mathbf{y}$ in the error layer: $\mathbf{e}=\mathbf{x}-\mathbf{y}$. The error is transformed by
the bottom-up (BU) transformation matrix $\mathbf{W}$ and forms the BU transformed error $\mathbf{s}$. BU
transformation maximizes BU information transfer in order to facilitate reconstruction.  BU transformed error is passed
to the internal representation layer through transformation matrix $\mathbf{N}$ (the role of this transformation shall
be discussed later) and is added to the internal representation $\mathbf{h}$ of the `hidden' (or model, or internal
representation) layer. The activity of the hidden layer is maintained by diagonal elements of recurrent matrix
$\mathbf{M}$, which -- beyond temporal integration -- can support category formation \cite{keri02categories} as well as
temporal prediction \cite{rao97dynamic} via its off-diagonal elements.

The network of Fig.~\ref{f:recnets}(B) works as follows. Off-diagonal elements of recurrent matrix $\mathbf{M}$ of the
hidden layer perform pattern completion \cite{lorincz02mystery} and temporal prediction \cite{rao97dynamic}. The newly
introduced layer denoted by $\mathbf{s}$ ($\mathbf{s} \in \mathbb{R}^r$) will be called the ICA layer. This ICA layer
plays a role in noise filtering. There are two different sets of afferents to the ICA layer: one is carrying the BU
transformed error, whereas the other carries information about the reconstructed input $\mathbf{y}$ via bottom-up
transformation $\mathbf{P}$, followed by a non-linearity that removes noise via thresholding. Thresholding is alike to
wavelet denoising \cite{mallat98wavelet}, with the exception that the filters are not necessarily wavelets but are
optimized for the input database experienced by the network. Optimization makes use of independent component analysis
(ICA)
\cite{jutten91blind,comon94independent,laheld94adaptive,bell95information,amari96new,karhunen97class,amari98natural}.
ICA makes the assumption that input is generated by statistically independent sources
\begin{eqnarray}\label{eq:ICA}
\mathbf{x} &=& \mathbf{C} \mathbf{r}+\nu \\
P(r_1, \ldots , r_r) &=& \prod_{k=1}^r P(r_k)
\end{eqnarray}
where $\nu \in \mathbb{R}^r$ denotes additive Gaussian noise, $\mathbf{r} \in \mathbb{R}^r$ denote the original sources
and $\mathbf{C} \in \mathbb{R}^{r \times r}$ is called the mixing matrix. The ICA algorithm intends to reproduce the
original sources and minimizes mutual information between ICA transformed components. The multiplication of vector
$\mathbf{r}$ by matrix $\mathbf{C}$, makes the identification of matrix $\mathbf{C}$ ill posed: The problem becomes
well posed, by constraining the variances of the searched components. For example, variances can be constrained to one.
In this case, minimization of mutual information is equivalent to maximizing the sum of the \textit{negentropies} (the
non-Gaussian character) of uncorrelated estimates \cite{hyvarinen99sparse}. This feature enables the local estimation
and local thresholding of noise components \cite{hyvarinen99sparse,hyvarinen99sparse2}, where locality means that noise
of each component of the ICA layer can be thresholded independently of the value of other ICA components. The method is
called sparse code shrinkage (SCS) and the process is referred to as sparsification.

Thus, $\mathbf{s}$ is the ICA transformed and sparsified reconstruction error. Note however, that sparsification
concerns the components of the BU transformed reconstructed input: high amplitude components of $\mathbf{Py}$ (BU
transformed reconstructed input) open the gates of components of the ICA layer and ICA transformed reconstruction error
can pass these open gates to correct the internal representation. Low amplitude components of $\mathbf{P}$ transformed
reconstructed input can not open the gates and corrections of these components are rejected, unless the correction
themselves are large enough to overcome the sparsification process.

Apart from sparsification, the reconstruction network is a linear network. In what follows, we shall denote this
property by the notations sign `$\sim$'. $A$ $\sim$ $B$ means that for a well tuned system and up to a scaling constant
(or scaling matrix) quantity $A$ is approximately equal to quantity $B$.

For a well tuned network and if matrix $\mathbf{M}$ performs temporal integration (i.e., if matrix $\mathbf{M}$ does
not perform temporal prediction), $\mathbf{s}\sim \mathbf{\dot{x}}$ by construction. If matrix $\mathbf{M}$ performs
temporal prediction and the network is perfectly tuned, then $\mathbf{s}\sim \mathbf{\ddot{x}}$. These quantities could
be processed and represented at higher layers. For the sake of simplicity of considerations, assume that matrix
$\mathbf{M}$ performs temporal integration and nothing else. Then we can approximate the output of layer $\mathbf{s}$
as a noise filtered linearly transformed (i.e., filtered and scaled) form of $\mathbf{\dot{x}}$.\footnote{Extension to
spatio-temporal pattern completion is straightforward. Redefining the state as the concatenation of $\mathbf{x}$ and
$\mathbf{\dot{x}}$ and speed as the concatenation of $\mathbf{\dot{x}}$ and $\mathbf{\ddot{x}}$ gives rise to the same
formalism that we are using here \cite{szepesvari97approximate}.} We shall further simplify the notation. The output of
the ICA layer will be denoted by $\sim \mathbf{\dot{x}}$, where $\sim \mathbf{\dot{x}}$ means `the scaled version' of
$\mathbf{\dot{x}}$. In a similar vein, internal representation $\mathbf{h} \sim \mathbf{y}$. Also, we shall use the
notation $\mathbf{y}\sim \mathbf{x}$, although the former is the noise filtered version of the latter.

Note that considerable reconstruction error can build up, e.g., by top-down influence. Considering longer temporal
scales, then  -- by construction -- $\mathbf{y}$ is equal to the temporal integral of the noise filtered reconstruction
error: $\mathbf{y} = \int \mathbf{e}_f dt$ where $\mathbf{e}_f$ denotes the noise filtered reconstruction error. This
feature will be exploited later.


\section{The joined model}
\label{s:model}

\subsection{Controlled reconstruction network}
\label{ss:control_}

\begin{figure}[h]
\centering
 \includegraphics[width=110mm]{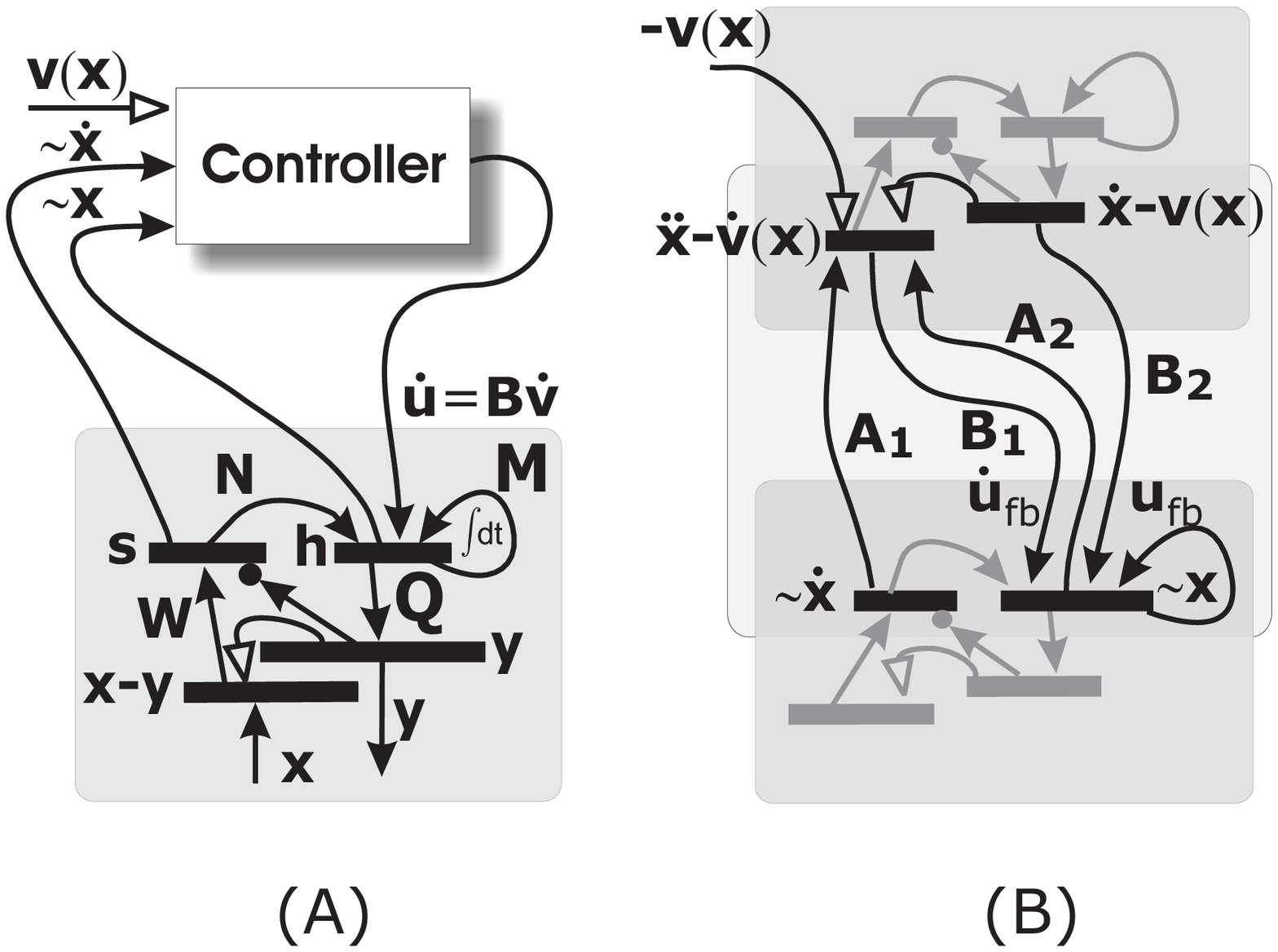}
 \caption{\textbf{Controlled networks}
\newline \noindent  \textit{Notation:} $\sim \mathbf{z}$: scales (approximately) linearly with
  $\mathbf{z}$.
  \newline \textbf{A:} Controller receives the state ($\sim \mathbf{x}$) and the speed ($\sim \mathbf{\dot{x}}$)
  from the reconstruction network under control as well as the desired speed ($\mathbf{v}(\mathbf{x})$) from
  somewhere else and forms the difference between them.
  \newline \textbf{B:} Reconstruction network acting as a controller. $\mathbf{A}_1$ and $\mathbf{A}_2$
  carries information about the state and the speed of the network under control, respectively. Difference between
  speed and desired speed is the input to the network. The noise filtered and reconstructed value of this difference
  appears at the reconstructed input. By construction, the reconstruction error layer holds -- the approximate --
  temporal   derivative of the reconstructed input. These differences are turned into feedback control vector
  $\mathbf{u}_{fb}$ and its temporal derivative $\mathbf{\dot{u}}_{fb}$  by means of transformations
  $\mathbf{B}_2$ and $\mathbf{B}_1$, respectively. The outputs of these transformation are temporally
  integrated at the internal representation of the lower network to form the components of the SDS controller.
  They add up here and provide almost perfect and stable control.} \label{f:controlled_rec_net}
\end{figure}

The reconstruction network can be controlled (Fig.~\ref{f:controlled_rec_net}(B)). Control adds extra contribution to
the hidden layer, namely $\mathbf{B}\mathbf{\dot{v}}$. The `dot' on $\mathbf{v}$ is the consequence of adding the
contribution to the internal representation where temporal integration occurs. In turn, the action of the controller is
equal to the temporally integrated value of the extra contribution, that is $\mathbf{B}\mathbf{v}$.

To achieve approximately perfect controlling, we shall make use of the SDS controller. The controller is made of
another (`higher') reconstruction network, which receives input from the network under controlled (the `lower'
network). This input is equal to the output of the ICA layer of the lower network. Control works by subtracting the
desired speed from the input of the higher network. That makes the input to the network equal to
$\mathbf{\dot{x}}-\mathbf{v}(\mathbf{x})$.\footnote{Note the negative sign of this difference that we shall discuss in
the next paragraph} By construction, (i) the input is noise filtered and reconstructed at the reconstructed input of
the network and (ii) apart from the noise content, the error layer approximates the temporal derivative of the
reconstructed input (Fig.~\ref{f:controlled_rec_net}(B)). These two differences undergo linear transformation and enter
the internal representation layer of the lower network, where -- by construction -- they add up and undergo temporal
integration. This is exactly what is needed for the SDS controller. Moreover, noise filtering of the reconstructed
input is of particular importance, because noise entering temporal integration seems to be the weakest point of the
controller \cite{szepesvari97approximate}.

Learning to control in the SDS scheme is relatively simple. Roughly speaking, learning is sufficient if the signs of
the components of the control vector and the domains where these components should not change sign have been
determined. This is why the negative negative sign of $\mathbf{\dot{x}}-\mathbf{v}(\mathbf{x})$ (instead of
$\mathbf{v}(\mathbf{x})-\mathbf{\dot{x}}$) does not count: signs of components of matrix $\mathbf{B}$ need to be
determined by learning. SDS warrants that if the signs of the control components are proper, then control will be
globally stable and approximately perfect. Mathematical details of \textit{sign-properness} can be found elsewhere
\cite{szepesvari97neurocontroller,szepesvari97approximate,lorincz_http_arxiv_CHF}. Numerical demonstrations using
coarsely tuned but sign-proper controllers have been provided in \cite{lorincz01ockham}.

The hierarchy is highly non-linear because of several reasons. The higher the reconstruction network in the hierarchy,
the higher the order of the dynamic properties of the environment that are learned and represented by it. Also, the
condition of robust control concerns the shattering of the space to sign-proper domains and the dynamic contribution of
the controller can be highly non-linear within sign-proper domains. Switching between domain is highly non-linear, too.
Lastly, sparsification is another source of non-linear processing.

Another note concerns the result of controlling. The SDS controller warrants that the desired quantity is closely
approximated under the condition that the system is controllable. Until this point control concerned a lower
reconstruction network, which may receive input from the environment (Fig.~\ref{f:controlled_rec_net}(B)). Conditions
of the SDS controller are not fulfilled unless control action also concerns this input, or if this input is zero. In
the former case, control of the environment needs to be included into the architecture. The latter case corresponds to
the absence of external inputs, such as pattern completion or dreaming.

The perfectly tuned architecture behaves as a bottom-up feedforward network, which is biased by top-down influence.
Consider the lower reconstruction network of Fig.~\ref{f:controlled_rec_net}(B). The bias will modify the internal
representation of the lower reconstruction network, which may or may not fit the input from the environment. If it
fits, then any reconstruction error disappears quickly. In case if it does not fit, large reconstruction error at the
reconstruction error layer may build up, but only a small portion of this error can pass the sparsification process at
the ICA layer. This is because sparsification is ruled by the internal representation generated reconstructed input.
That is, information that matches the \textit{context} of the higher reconstruction network will be able to pass
sparsification, whereas other information will be \textit{filtered out}, or attenuated by the bottom-up sparsification
process. In turn, top-down influence `paves the way' of some of the components. The process of filtering out certain
components of information and paving the way of the others can be seen as \textit{attention is paid} to the latter
components.

\subsection{Closing the loops of the hierarchy} \label{ss:closing}

\begin{figure}
\centering
 \includegraphics[width=80mm]{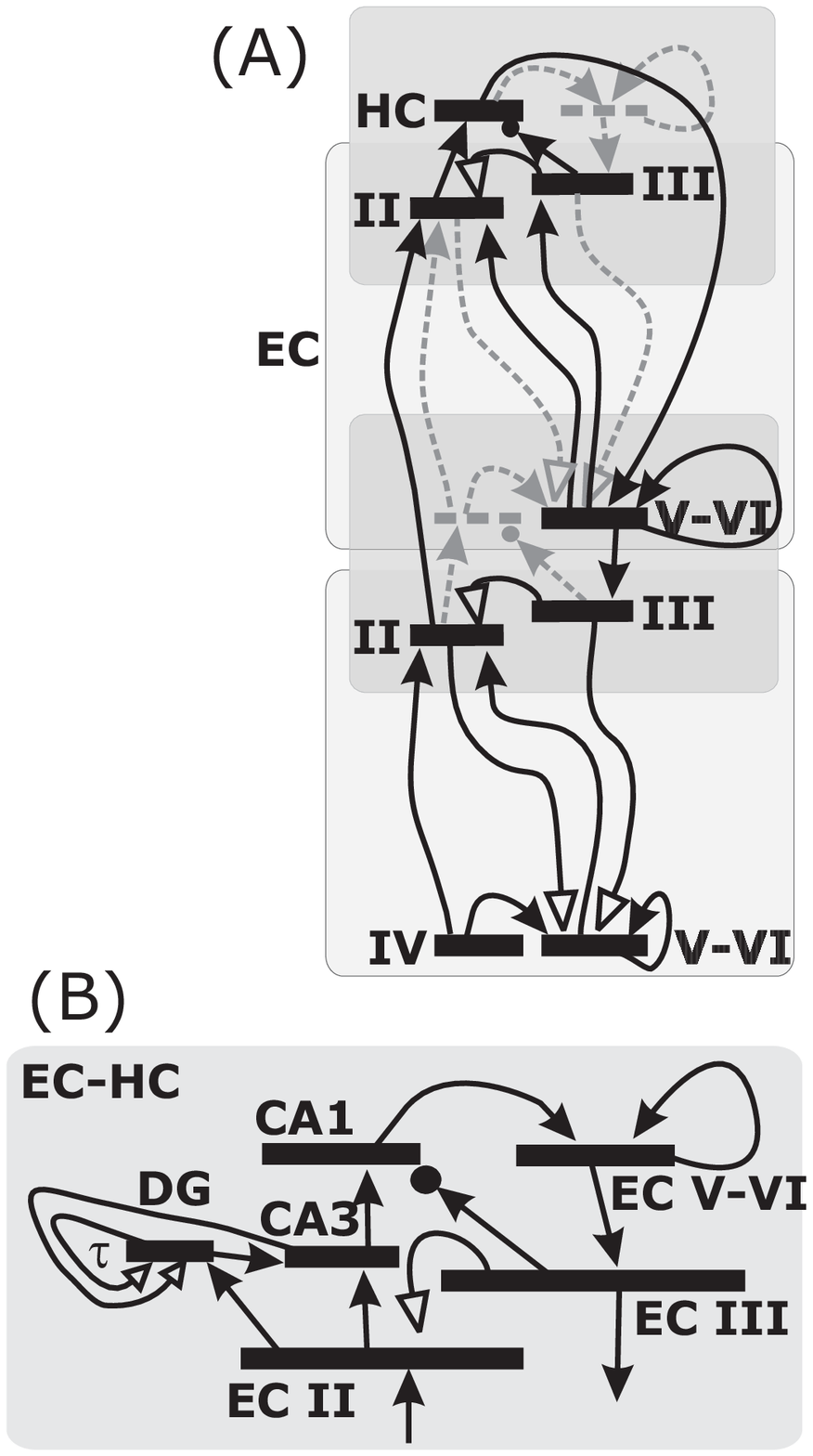}
  \caption{\textbf{The top of the hierarchy}
  \newline \noindent  \textbf{A:} HC: hippocampus. II-III: superficial layers. IV: granular layer. V-VI: deep layers.
  The top turns information flow from bottom-up direction to top-down direction: HC acts on the internal representation
  of a lower reconstruction network. HC together with the layers encompassed by the grey box indexed by EC,
  i.e., the entorhinal cortex, which has no granular layer, form a reconstruction network.
 \newline \textbf{B:} Mapping of the top to the EC-HC loop.
  ICA is made in two steps: (i) whitening (CA3 subfield of HC)
  and (ii) separation (CA1 subfield of HC). Blind source deconvolution, the putative role
  of the dentate gyrus, removes temporal correlations.
  Reconstruction error is computed at EC layer II.
  Reconstructed input: EC layer III. Internal (hidden) representation: EC layers V and
  VI. Temporal integration (persistent activities) and possibly pattern completion
  is the putative role of the recurrent collaterals of EC layers V and VI.
  EC afferents of area CA1: denoising.}\label{f:top}
\end{figure}

The top of the hierarchy has a special role. At the top, the sensory bottom-up information flow should be turned into
context like top-down information, which can enforce, shape, or influence (say, control) lower representations. This
reversal of the direction of information flow becomes possible if the reconstruction error feeds the internal
representation of a \textit{lower} reconstruction network. In this case, we shall have a network with a reconstruction
error layer and an internal representation, which when put together, make up a reconstruction network. On the other
hand, we have also a controller, because error at a higher level guides the internal representation of a lower level.
The idea is illustrated in Fig.~\ref{f:top}(A). The figure uses the same gray coding that
Fig.~\ref{f:controlled_rec_net}: dark and light gray areas correspond to reconstruction and control networks,
respectively. There are two dark and two light areas. All of the components are depicted, but some of them are denoted
by lighter dashed lines. These connectivity structures and neural layers are missing at the top. Notation corresponds
to the cortical layers. Layers $\mathbf{x}$, $\mathbf{y}$, $\mathbf{s}$, and $\mathbf{h}$ represent superficial layers
II, III, granular layer IV, and deep layers V-VI, respectively. The highest layer is denoted by HC, the short for
hippocampus. The upper control layer is denoted by EC, the short for the entorhinal cortex.

The following differences are to be noted. First, there is no top-down connection from EC layers II and III to EC the
deep layers. In turn, these layers can not exert control action onto the corresponding internal representation.
Instead, the HC is in the position to exert control action: it receives the reconstruction error from the lower layer
and acts upon the hidden representation of that reconstruction network. On the other hand, consider the connectivity of
the deep layers of the EC, the superficial layers of the EC and the HC. These structures, together, form a
reconstruction network. It can be easily verified by shifting the deep layers of the EC next to the HC in
Fig.~\ref{f:top}(A). In turn, the HC and its environment has a double role, it is a control network and a
reconstruction network. The control network acts upon a lower internal representation that reverses the bottom-up flow
of sensory information into top-town control making use of the context of higher reconstruction networks. Using the
concepts of the controller hierarchy, higher order dynamical information corrects lower order approximations.

Note that information flow from the lowest layers ($\mathbf{s}$ and $\mathbf{h}$ passes through the ladder of two
reconstruction error layers (superficial layer II). This feature can not be explained within the framework of the
model. It allows us to pin-point to the limitations of our modelling efforts. The functional model can not explain the
recurrent collaterals of the superficial layers, a prominent structure of neocortical regions. This connectivity is
thought as the extension of the associative cortices (see, e.g., \cite{diamond79subdivision} and references therein).
Our model suggests that two of such layers can be seen as a single but larger layer, which is in agreement with the
functionality suggested by Diamond.

Detailed description of the HC has been provided elsewhere \cite{lorincz00parahippocampal,lorincz02mystery} and will be
reviewed in the next section.

\section{Discussion} \label{s:discussion}

Our first note concerns Adaptive Resonance Theory (ART) pioneered by Grossberg and colleagues
\cite{grossberg80how,carpenter87massively,raizada03towards}. ART proposes that sensory processing is two-folded: It is
made of bottom-up filtering as well as of top-down template matching. The underlying mechanism of ART -- namely, the
resonant circuitry -- differs from the two-folded comparator function embodied by our architecture. Mapping to the
neurobiological substrate \cite{grossberg93normal,raizada03towards} is, in turn, different from ours that we shall
present below.

Our modelling efforts have certain particular properties: We have started from functional hypothesis about the
importance of control and noise filtering and have introduced a hierarchical system, which should be capable to do
both. At each step, mathematical tools were used to restrict our freedom. Possible solutions, which were untractable
from the point of view of stability and noise filtering, or, alternatively, which could not be mapped to the anatomical
structure, or did not fit known physiological properties and known results of computational neuroscience were dropped
\cite{lorincz00isthehippocampus}. We call this function based, structure constrained and mathematics supported effort,
Ockham's modelling \cite{lorincz01ockham,lorincz02modeling}. ART does the same. Our comparator model is different,
because it starts from control principles control principles and assumes the universality of the comparator hypothesis.
For a control network, comparison between desired and experienced quantities seems reasonable.

The model offers falsifying predictions; i.e., predictions which could constrain or defeat the model. These predictions
will be listed below. First, we shall provide the mapping to the substrate, a most crucial constraint for us. The
mapping of the reconstruction network has been elaborated before \cite{lorincz02mystery}. It is reviewed here for the
sake of completeness.

\subsection{Matching the anatomy} \label{ss:matching}

\begin{figure}[h!]
 \centering
 \includegraphics[width=80mm]{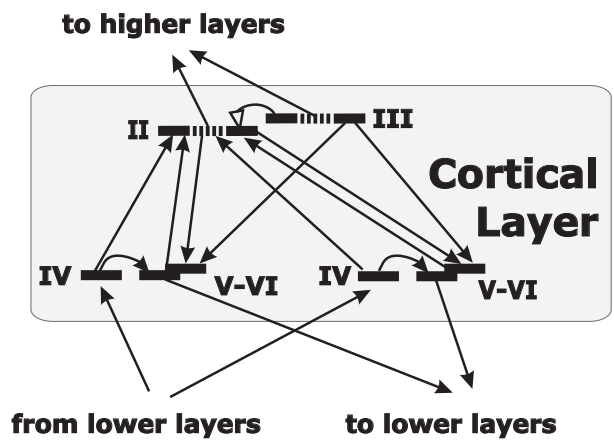}
  \caption{\textbf{Neocortical layer}
  }\label{f:neocortex}
\end{figure}

The neocortex is made of six sub-layers (Fig.~\ref{f:neocortex}). The figure depicts the most prominent connections
between these sub-layers \cite{lund88anatomical}. Input typically arrives at layer IV. Layer IV neurons send messages
to layer II and layer III (not shown). Furthermore, layer IV neurons send messages also to layer VI. Superficial
neurons provide output down to layer V and VI. There are connections between neurons of layer V and layer VI. Neurons
of layer II and III are also strongly connected. Layer V provide feedback to layers II and III. The main output to
higher cortical layers emerges from layers II and III. The main feedback to lower layers is provided by layer V. (For a
review see, e.g., \cite{callaway99visual}.)

The theoretical model and the anatomical structure can be matched by assuming that reconstruction networks are laid
\textit{between} neocortical layers as it was denoted by the dark gray areas in Fig.~\ref{f:top}(A)
\cite{lorincz02mystery}. According to this figure, superficial layers of the lower cortical layer \textit{and} deep
layers of the higher cortical layer form \textit{one} functional unit, the reconstruction network.

The novel anatomical suggestion of this paper is the mapping of a robust controller to the cortical layers (denoted by
light gray boxes of Fig.~\ref{f:top}(A)). The reconstruction error and the noise filtered temporal integral of this
error, that is the reconstructed input exerts control action on lower reconstruction networks that corresponds to the
information sent from superficial layers to deep layers. Experienced quantities correspond to the information
propagating into the other direction, i.e., experienced quantities are sent to superficial layers by layers IV and V.

Mapping of the top to the EC-HC loop is shown in Fig~\ref{f:top}(B). Independent component analysis is executed in two
steps: (i) whitening (CA3 subfield of HC) and (ii) separation (CA1 subfield of HC). Reconstruction error and
reconstructed input are computed in EC layer II and EC layer III, respectively. Internal (hidden) representation is
encompassed by EC layers V and VI. These deep layers perform the pattern completion task. The EC afferents of the CA1
subfield can release the gates of the CA1 outputs and denoised signals can pass. The recurrent collaterals of the CA3
subfield replay learned sequences. Blind source deconvolution (BSD) is the putative role of the dentate gyrus. BSD
removes temporal correlations cumulated by temporal integration of not-yet tuned lower reconstruction networks. BSD is
necessary for proper ICA analysis. However, the number of neurons of a BSD structure can be very large. It is assumed
that this computation can be afforded only at the top of the hierarchy. In turn, the dentate gyrus plays an unique role
in our model. The assumption gains further support from top-down controlling: control can accelerate reconstruction
\cite{lorincz98forming}. In turn, BSD at lower networks may not be necessary. This point deserves further
investigations. More details can be found elsewhere \cite{lorincz00parahippocampal}.

\subsection{Predictions of the model} \label{ss:disc_attract}

L{\H o}rincz and colleagues have shown previously that some memory effects, such as repetition suppression and priming
\cite{lorincz02relating} as well as the particular properties found for Alzheimer patients in a classic prototype
learning paradigm (see, e.g., \cite{knowlton99what} and references therein about `9 dot' experiments) can be explained
by the reconstruction network model \cite{keri02categories}.

A crucial prediction of our model is that temporal convolutions should be removed before ICA occurs. Given that ICA is
the putative role of the CA3 and CA1 subfields, only the dentate gyrus can be responsible for this task. In turn, long
and tunable delay lines should exist in the dentate gyrus. This prediction has been reinforced recently
\cite{henze02single}.

Another falsifying prediction of the model concerns the internal representation layer, which has to maintain its own
activities in order to enable additive corrections and temporal integration. Persistent activities in the deep layers
but not in the superficial layers of the EC have been found experimentally \cite{egorov02graded}.

An intriguing and falsifying prediction of the joined model, alike to its previous versions
\cite{lorincz00parahippocampal}, is that top-down connections of reconstruction networks can be interpreted as
long-term memories (LTMs), because these connections are responsible for the relaxed activities of the hidden layers.
Given our mapping, the LTM corresponds to feedback connections between neocortical areas. These connections are
generally more numerous than the feedforward connections between the same areas but the activity flow along these
connections is relatively low and suggests a weak functional role (see, e.g., \cite{callaway99visual} and references
therein). This apparent discrepancy may be resolved by noting that different interpretations may coexist in the brain
as it has been made evident in the animal experiments on binocular rivalry (see, e.g., \cite{leopold99multistable} and
references therein) and in experiments with several possible visual interpretations (see, e.g.,
\cite{leopold03visual,parker03neuronal} and the cited references). If reconstruction concerns a single interpretation
then feedback activity flow should be \textit{small}. This possibility can not be excluded because of the following
reasons. There are evidences that activities in V4 (responsible for conscious detection of colors) and V5 (responsible
for conscious detection of fast motion) in the monkey are uncorrelated. According to the arguments put forth by Zeki
\cite{zeki03disunity}, uncorrelated activities indicate that conscious experiences propagate downwards along parallel
channels. Moreover, the conscious binding of the result of the individual conscious experiences seems to be delayed
\cite{bartels02temporal}. In turn, it is possible that only one interpretation is communicated downwards at a time.
Another point concerns the suggested function, that the internal representation is under top-down control. This
controlled representation will then propagate downwards to form the reconstructed input and also directly to real
control networks \cite{diamond79subdivision}. Given the long delays of processing, a well tuned control system should
not interact oftentimes. In turn, the assumed function involves relatively sparse information flow.

It is important to note that in the original \textit{control} model of the EC-HC loop \cite{lorincz98forming} the
dentate gyrus was suggested as the source of temporal integration. Only the making of the reconstruction network model
\cite{lorincz00parahippocampal,chrobak00physiological} revealed that temporal integration has to be executed at the
hidden layer and not at the dentate gyrus, whereas temporal convolutions produced by temporal integrations accomplished
in lower networks can be removed by means of the BSD algorithm at the dentate gyrus.

According to recent measurements, awareness and attention needs to be distinguished (for an excellent review, see
\cite{lamme03whyvisual}). Attention increases neuronal activities responsible for the processing of the attended
stimuli \cite{desimone95neural}. Most probably, endogenous attention facilitates the pathways that should be used by
the attended stimuli \cite{egeth97visual}. In our model, facilitation can manifest itself through control action
\textit{within} cortical layers. On the other hand, awareness involves recurrent interactions between areas and can be
suppressed by backward masking (see \cite{lamme03whyvisual} and references therein). This recurrent interaction
required for awareness is our candidate function of the feedback connections between cortical areas.

Another note concerns top-down pattern completion: The controller network combines bottom-up information from different
columns, areas and modalities and develops the \textit{context} for lower level internal representation. Control action
then corresponds to context based pattern completion. The efficiency of strong top-down control, such as overwriting,
has been the subject of computer studies \cite{lorincz02mystery}.

The merging of the two kinds of comparator networks solves the noise sensitivity of the controller. This noise
filtering is fast and optimal, and it is an emerging property in the joined architecture.

According to the model, the EC deep layer to EC superficial layer synapses form the long-term memory of the EC-HC loop.
On the other hand, the HC plays a particular, though not unique role. The HC is the top comparator that turns
reconstruction error into control signal. This control signal excite the neurons of the EC deep layers. This is a
unique position to encode inputs into the synapses between EC superficial and deep layers by Hebbian means. Similar
roles can be played by all top-down control signals at other levels. These control signals target the deep layers and
similarly to the HC output, they may enable (or facilitate) Hebbian learning of the LTM. Moreover, the control signal
may propagate from the higher reconstruction networks to lower ones and encoding at higher reconstruction networks may
influence encoding at lower ones, too.

Our model, alike to its previous versions \cite{lorincz98forming,lorincz00parahippocampal,lorincz02mystery} is neither
a model for episodic learning, nor a model for incremental learning and does not fit such traditional distinctions
(see, e.g., \cite{gluck03computational}). On the one hand, when information maximization is not modulated by behavioral
relevance, the model is an incremental model engaged in the maximization of information transfer, noise filtering and
pattern completion using information theoretic algorithms in Hebbian forms \cite{lorincz02mystery}. On the other hand,
the controller is a top-down tool, which can facilitate learning and could make learning instantaneous, if behavioral
relevance requires: For a given input, and by activating a unit of the hidden layer, Hebbian learning will make that
unit of the hidden layer to represent (encode) the actual input by means of its top-down synapses. Such mechanism can
shortcut statistical analysis and may imprint the input into the internal representation. According to our model, the
increased learning rate in deep layer afferents of the superficial layers corresponds to the supervisory instruction
for the activated deep layer unit(s): Remember to the actual input!

The joined model offered no role for the recurrent collaterals of the superficial layers. We believe, that our
continuous model can not uncover the role of these connectivity structures. Another missing feature of the neocortical
structure is its columnar organization. The continuous comparator model does not seem to offer any clue here.

Finally, we note that the forming of invariant place cells from retinal input irrespective of the motion of eyes, head
and body and their learned and optimized joined or disjoined motion patterns corresponds to a plant of very high order.
Our rate code model justifies that invariant representations of place cells, the behaviorally important components of
problem solving in mazes, are represented the hippocampus in rats  and that the information of different modalities are
associated here. (For a review, see, \cite{redish99beyond}).

\section{Conclusions} \label{s:conc}

There is a large body of experimental data supporting the idea that attention shapes (influences, controls) perception.
For excellent reviews, see, e.g., \cite{duncan99attention,posner00attention,laberge00networks} and references therein.
We have presented a unified model that optimizes bottom-up information transfer and filters (attenuates, prohibits) the
propagation of structureless noise. The model also influences top-down processes, by comparing desired and experienced
parameters in sensory information processing. The unification of the control model \cite{lorincz98forming} and the
reconstruction architecture model \cite{lorincz00parahippocampal} leads to falsifying predictions. Some of those
predictions have gained experimental support recently. For example, the model predicts persistent activities in the
deep layers of the entorhinal cortex that have been found in the experiments \cite{egorov02graded}. Another falsifying
prediction, that the circuitry of the dentate gyrus should support long delays, has also been reinforced
\cite{henze02single}. A most intriguing prediction of the model is that the long-term memory of neocortical visual
areas corresponds to the feedback connections between areas. These feedback connections are more numerous than the
bottom-up connections between areas. However, these feedback connections are relatively quiet. The model allowed us to
distinguish between attention and awareness, two delicate and intertwined concepts. Known features of awareness allowed
us to argue about the relative quietness of feedback connections between areas: there should be only one available
representation for awareness, whereas multiple interpretations should coexist in representations not directly related
to awareness. We have argued that this interpretation fits recent physiological findings.

\section*{Acknowledgements}
\label{s:acknow}

This work was partially supported by Hungarian National Science
Foundation (Grant No. OTKA T-32487). Special thanks are due to
Gy\"orgy Buzs\'aki for his enlightening and continuous support
during the long-course of our model construction. Careful reading
of the manuscript and helpful suggestions are gratefully
acknowledged to Gy\"orgy H\'ev\'izi and to G\'abor Szirtes. I
should thank the enlightening critical notes of Peter Dayan that
have helped me in improving the presentation.

\clearpage \newpage \bibliographystyle{apalike}


\end{document}